\documentclass[prb, reprint, superscriptaddress]{revtex4-2}

\usepackage{graphicx}
\usepackage{dcolumn}
\usepackage{bm}
\usepackage{xcolor}

\usepackage{listings} 
\usepackage{enumerate} 

\definecolor{codegreen}{rgb}{0,0.6,0}
\definecolor{codegray}{rgb}{0.5,0.5,0.5}
\definecolor{codepurple}{rgb}{0.58,0,0.82}
\definecolor{backcolour}{rgb}{0.95,0.95,0.92}

\lstdefinestyle{mystyle}{
    backgroundcolor=\color{backcolour},   
    commentstyle=\color{codegreen},
    keywordstyle=\color{magenta},
    numberstyle=\tiny\color{codegray},
    stringstyle=\color{codepurple},
    basicstyle=\ttfamily\footnotesize,
    breakatwhitespace=false,         
    breaklines=true,                 
    captionpos=b,                    
    keepspaces=true,                 
    numbers=left,                    
    numbersep=5pt,                  
    showspaces=false,                
    showstringspaces=false,
    showtabs=false,                  
    tabsize=2
}
\usepackage{amsmath}
\graphicspath{{./pics/}}

\usepackage{mathrsfs}  
\usepackage{float} 

\usepackage[colorlinks = true, linkcolor=blue, citecolor=blue, urlcolor  = blue]{hyperref}

\def\equationautorefname~#1\null{Eq.~(#1)\null}

\usepackage{orcidlink}

\begin{document}

\title{Predicting Interacting Green's Functions with Neural Networks}

\author{Egor Agapov\,\orcidlink{0000-0002-5000-1277}}
\email{egor.agapov@tum.de}
\author{Oriol Bertomeu\,\orcidlink{0009-0002-8131-0555}} 
\email{oriol.bertomeu@tum.de}
\author{Andrés Carballo\,\orcidlink{0009-0009-4018-3319}} 
\email{andres.carballo-santana@tum.de}
\affiliation{Ludwig-Maximilians-Universität München, Munich, Germany}
\affiliation{School of Natural Sciences, Technical University of Munich, Garching, Germany}

\author{Christian B.~Mendl\,\orcidlink{0000-0002-6386-0230}}
\email{christian.mendl@tum.de}
\affiliation{Department of Computer Science, Technical University of Munich, Garching, Germany}
\affiliation{Institute for Advanced Study, Technical University of Munich, Garching, Germany}


\author{Aaron Sander\,\orcidlink{0009-0007-9166-6113
}}
\email{aaron.sander@tum.de}
\affiliation{Chair for Design Automation, Technical University of Munich, Munich, Germany}

\date{\today}

\begin{abstract}
Strongly correlated materials exhibit complex electronic phenomena that are challenging to capture with traditional theoretical methods, yet understanding these systems is crucial for discovering new quantum materials. Addressing the computational bottlenecks in studying such systems, we present a proof-of-concept machine learning-based approach to accelerate Dynamical Mean Field Theory (DMFT) calculations. Our method predicts interacting Green's functions on arbitrary two-dimensional lattices using a two-step ML framework. First, an autoencoder-based network learns and generates physically plausible band structures of materials, providing diverse training data. Next, a dense neural network predicts interacting Green’s functions of these physically-possible band structures, expressed in the basis of Legendre polynomials. We demonstrate that this architecture can serve as a substitute for the computationally demanding quantum impurity solver in DMFT, significantly reducing computational cost while maintaining accuracy. This approach offers a scalable pathway to accelerate simulations of strongly correlated systems and lays the groundwork for future extensions to multi-band systems.
\end{abstract}

\maketitle

\section{Introduction}
In the study of condensed matter physics and materials science, 
understanding the electronic properties of strongly correlated materials presents a significant challenge. 
Traditional methods, such as Density Functional Theory (DFT), have proven immensely successful in describing weakly correlated systems, 
where the interaction between electrons can be treated as a relatively small perturbation \cite{PhysRev.136.B864, PhysRev.140.A1133}. 
However, when it comes to strongly correlated systems, where electron-electron interactions dominate, 
these conventional approaches often fail to capture the essential physics, such as localization of electrons, time-dependent fluctuation effects, and quasiparticle renormalization.  
Dynamical Mean-Field Theory (DMFT) can successfully capture the physics in these situations \cite{DMFTreview}. 
Nevertheless, implementing the algorithm is computationally demanding since the time and computational cost for the Monte Carlo method to converge grows with interaction strength.
In our work, we aim to use machine learning techniques to speed up the process by skipping this Monte Carlo process using machine learning methods.

Machine learning (ML) has been applied successfully across various challenges in condensed matter physics. For example, ML techniques have been used to refine approximate solutions of the Anderson impurity model (AIM) to match exact results \cite{weber_datadriven_2021}, predict AIM spectral functions on Bethe lattices \cite{sturm_good_2021}, and extract spectral functions directly from densities of states \cite{ren_good_2021}. Furthermore, ML has been explored for predicting Green’s functions in various representations \cite{arsenault_good_2014} and for accelerating Continuous Time Quantum Monte Carlo (CTQMC) routines \cite{song_accelerated_2019, lee_continuous_2019}. Building on these advancements, this work introduces a novel machine learning-based method designed to accelerate CTQMC calculations on arbitrary two-dimensional lattices. By leveraging neural networks to predict interacting Green's functions, our approach provides a scalable framework for integrating ML into the DMFT workflow, addressing key computational bottlenecks and enabling faster exploration of strongly correlated systems.

\begin{figure}
    \centering
    \includegraphics[width=\linewidth]{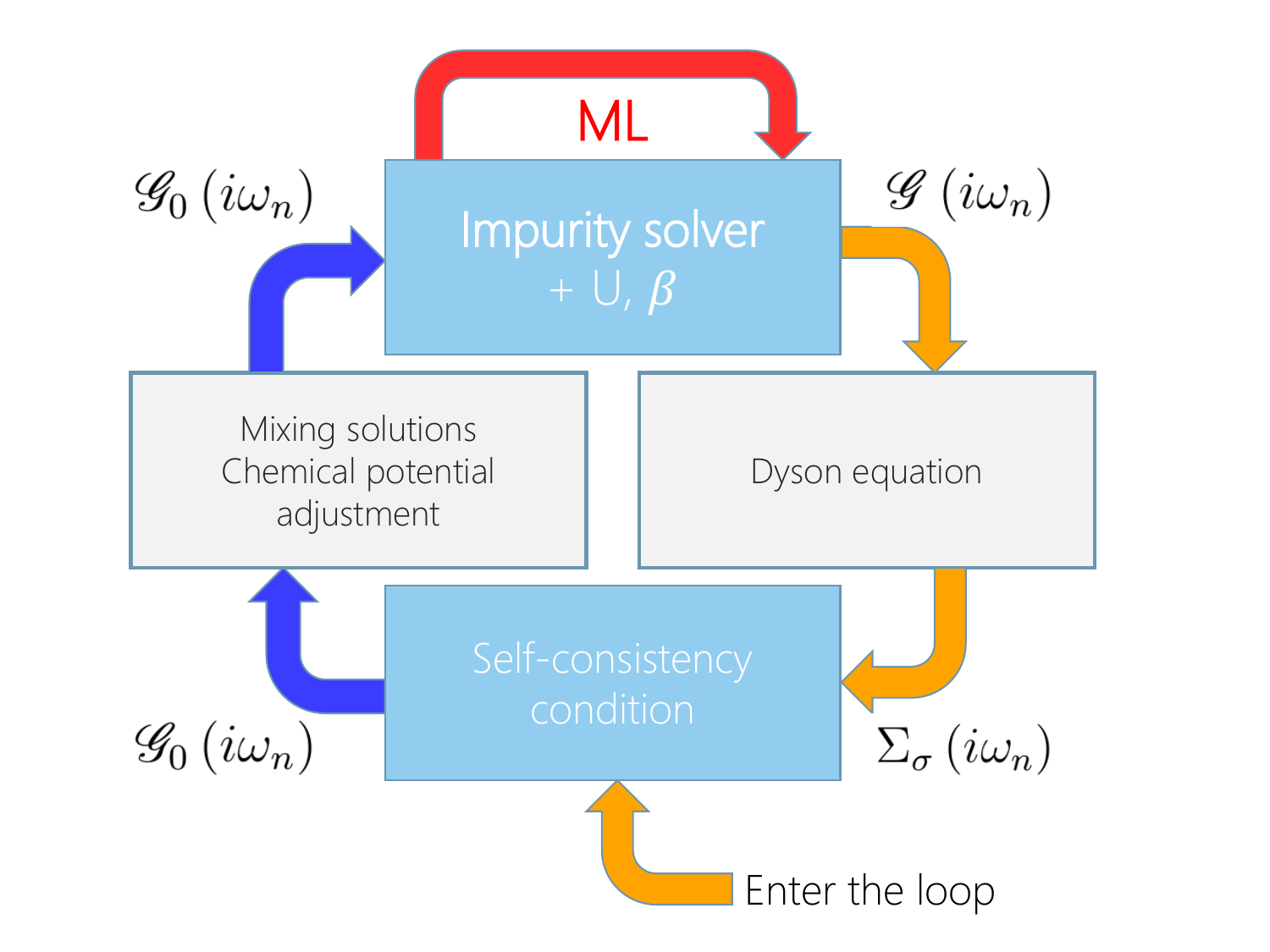}
    \caption{Schematic overview of the DMFT algorithm. 
    Here $\mathscr{G}_{0}$ is the non-interacting Green function, 
    $\mathscr{G}$ is the interacting one and $\Sigma_{\sigma}$ is the self-energy.}
    \label{fig:DMFTalgo}
\end{figure}  

\section{Preliminaries}
\subsection{Anderson Impurity Model}
DMFT simplifies the study of strongly correlated systems by mapping the original lattice problem onto a single impurity site embedded in an electron bath. This transformation reduces a complex many-body problem into a more manageable single-body problem, where the medium surrounding the impurity can be dynamically adjusted to capture the influence of the lattice. The core principle of this mapping is the self-consistency condition, which iteratively ensures that the impurity model accurately reflects the properties of the original lattice system.

The impurity problem in DMFT is typically described by the Anderson impurity model, a widely-used framework for modeling the interactions between electrons at the impurity site and the surrounding medium \cite{PhysRev.124.41}. While the general form of this model extends to multi-band cases, this paper focuses on the single-orbital, single-site Anderson impurity model with Hubbard interactions, which provides a foundational setting for developing and testing our machine learning-based approach.
The action of this model can be written as 
\[
    S = S_0 + S_U
\]
such that 
\begin{subequations}
    \begin{align}
        S_0 & = -\sum_\sigma \iint_0^\beta d \tau d \tau^{\prime} d_\sigma^{\dagger}(\tau) \mathscr{G}_\sigma^0\left(\tau-\tau^{\prime}\right)^{-1} d_\sigma\left(\tau^{\prime}\right), \\
        S_U & = U \int_0^\beta d \tau \ n_{\uparrow}(\tau) n_{\downarrow}(\tau),
    \end{align}
\end{subequations}
where $\beta$ is the inverse temperature, $\mathscr{G}_\sigma^0$ is the non-interacting Green's function, $d_{\sigma}$ are site operators, and $U$ is the Hubbard interaction at the impurity site.  

The self-consistency condition, central to DMFT, ensures that the impurity Green’s function matches the local Green’s function of the lattice. Mathematically, this is expressed as:
\begin{equation}
    \label{SC}
    \mathscr{G}_\sigma\left(i \omega_n\right) =\sum_{\boldsymbol{k}} G^{\text{loc}}_{\sigma}\left(i \omega_n, \boldsymbol{k}\right),
\end{equation}
where \( i \omega_n \) are Matsubara frequencies, and the summation is over all momentum states \( \boldsymbol{k} \) in the Brillouin zone.

Solving the impurity model yields local quantities, such as the Green’s function and the self-energy, which describe the electronic interactions within the system. These quantities are then used to update the effective medium, thereby modifying the impurity model itself (see \autoref{fig:DMFTalgo}). This iterative process is repeated until self-consistency is achieved, ensuring convergence of the impurity and lattice properties.

\subsection{Existing DMFT Solvers}
Dynamical Mean-Field Theory (DMFT) typically relies on Quantum Monte Carlo (QMC) solvers, such as continuous-time QMC (CTQMC), to compute the local Green’s function \( G^{\text{loc}}_{\sigma}\left(i \omega_n\right) \). The quality and efficiency of these methods depend heavily on the number of Monte Carlo iterations, with computations often taking hours to days for multi-band systems.
For QMC solvers, such as the Continuous-Time QMC (CTQMC), the computational time scales with both the interaction strength \( U \) and the inverse temperature \( \beta \)~\cite{DMFTreview}. In the strong-coupling regime, the time required to sample configurations and compute determinants increases due to the larger number of diagrams at higher \( U \). Similarly, lower temperatures (larger \( \beta \)) necessitate the inclusion of more Matsubara frequencies in the calculations, further increasing complexity.

In the Continuous-Time Hybridization-Expansion (CT-HYB) solver, the weights for the Monte Carlo Markov chain are expressed as:
\begin{equation}
    \begin{split}
        w\left(K,\left\{\alpha_j, \alpha_j^{\prime}, \tau_j, \tau_j^{\prime}\right\}\right) & \equiv 
        \operatorname{det}_{1 \leq i, j \leq K}\left[M^{-1}\right]_{i j} \\ 
        & \cdot \operatorname{Tr} \Bigl( \mathcal{T} e^{-\beta H_{\mathrm{loc}}} \prod_{i=1}^K c_{\alpha_i}^{\dagger}\left(\tau_i\right) c_{\alpha_i^{\prime}}(\tau_i^{\prime}) \Bigr),
    \end{split}
\end{equation}
where \( M \) is the determinant of the hybridization terms, \( H_{\mathrm{loc}} \) is the local interaction Hamiltonian, \( \hat{c}_{\alpha}(\tau) \) represents the impurity operator for orbital \( \alpha \) at imaginary time \( \tau \), and \( K \) denotes the perturbation order.

\begin{figure}
\centering
\includegraphics[width=0.99\linewidth]{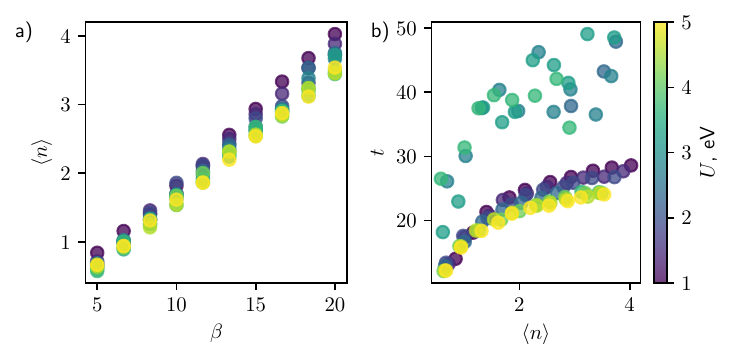}
\caption{(a) Dependence of the average perturbation order on \( \beta \), for various interaction strengths \( U \). (b) Required time in seconds for calculating the dependence on the average perturbation order. In subfigure (b), \( \beta \) grows with \( \langle n \rangle \).}
\label{fig:scaling}
\end{figure}

At low values of \( \beta \), the computational cost is dominated by the evaluation of the trace term, which scales logarithmically with \( K \). For sufficiently large \( \beta \), however, the determinant evaluation, scaling as \( K^3 \), becomes the dominant factor. In the case of single-orbital systems, this regime is typically not reached (see Fig.~\ref{fig:scaling}). Additionally, the computational cost’s dependence on \( U \) is reduced in the CT-HYB algorithm. By contrast, in the Continuous-Time Interaction-Expansion (CT-INT) solver, the average perturbation order scales as \( \beta U \), making it less efficient in strongly correlated regimes. In the Continuous-Time Auxiliary Field approach, the perturbation order scales even more steeply, as \( \beta U^2 \).

The quantum impurity solver represents the most computationally demanding step in the DMFT self-consistency loop, while the remaining steps require negligible computational resources. This bottleneck underscores the potential of ML-based approaches to significantly reduce the computational burden of DMFT by replacing or accelerating the impurity solver. In particular, we seekt to use ML techniques to accelerate this process by predicting \( G^{\text{loc}}_{\sigma}\left(i \omega_n\right) \) directly from the non-interacting Green’s function \( \mathscr{G}^{0}_{ \sigma} \) at each DMFT step.

\section{Neural Network-based Impurity Solver}
This section introduces our machine learning approach for accelerating the impurity solver in DMFT. We describe the architecture and training process of the \textbf{primary neural network}, which predicts the interacting Green’s function from a given non-interacting Green’s function. To generate the necessary training data, we develop an autoencoder-based \textbf{auxiliary neural network} that learns the band structures of known materials and generates physically plausible two-dimensional band structures.

The trained decoder of the auxiliary network serves as a generative model, providing diverse band structures from which non-interacting Green’s functions can be computed. These Green’s functions, combined with randomly sampled Hubbard interaction parameters \( U \) and inverse temperatures \( \beta \), form the dataset used to train the primary neural network. This approach ensures that the training data captures the complexity and diversity of systems relevant to DMFT, enabling accurate and efficient predictions of interacting Green’s functions.

\subsection{Primary Neural Network Architecture}
\begin{figure}
\centering
\includegraphics[width=0.99\linewidth]{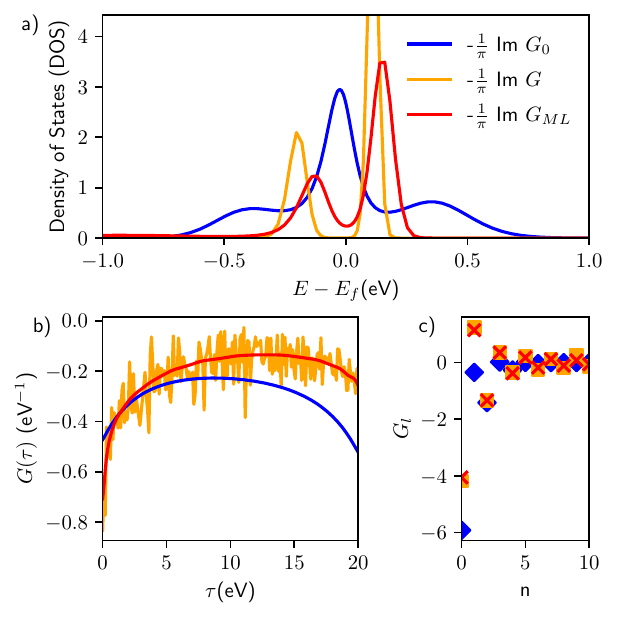}
\caption{Examples of (a) the density of states obtained from Green’s functions, (b) Green’s functions in the \( \tau \)-basis, and (c) Green’s functions in the Legendre representation. Comparisons are made for non-interacting Green’s functions (\( G_0 \)), interacting Green’s functions (\( G \)), and predicted Green’s functions (\( G_{\text{ML}} \)) for \( U = 4 \, \mathrm{eV} \) and \( \beta = 40 \, \mathrm{eV}^{-1} \).}
\label{fig:dos}
\end{figure}

The primary neural network predicts the interacting Green’s function from its non-interacting counterpart. Starting with a random band structure, we compute the non-interacting Matsubara Green’s function \( \mathscr{G}_{0}(i \omega_n) \) and its imaginary-time representation \( \mathscr{G}_{0}(\tau) \) using a Fourier transform. These Green’s functions are defined on the interval \( [0, \beta] \), which we rescale to \( [0, 1] \) to facilitate numerical handling. The Green’s functions are then expressed in terms of Legendre polynomials~\cite{boehnke_orthogonal_2011}, as follows:
\begin{subequations}
    \begin{align}
        G(\tau) &= \frac{1}{\beta} \sum_{n \leq 0} \sqrt{2n+1} \, P_n[x(\tau)] G_n, \\
        G_n &= \sqrt{2n+1} \int_0^\beta d\tau \, P_n[x(\tau)] G(\tau),
    \end{align}
\end{subequations}
where \( P_n \) are the Legendre polynomials (illustrated in \autoref{fig:dos}), and \( x(\tau) = \frac{2\tau}{\beta} - 1 \). The Legendre coefficients \( G_n \) compactly represent the Green’s functions and serve as the input features (non-interacting Green’s function) and target outputs (interacting Green’s function) for the primary neural network.

The primary neural network consists of a sequential architecture with four fully connected (dense) layers containing 64, 128, and 64 neurons, respectively, and employs ReLU activation functions for non-linearity (\autoref{fig:arch} (b)). The model is trained to minimize the Mean-Squared Error (MSE) loss, defined as:
\begin{equation}
\mathrm{MSE} = \frac{1}{N_{\ell}} \sum_{\ell=1}^{N_{\ell}} \left(G_{\ell} - \hat{G}_{\ell} \right)^2,
\end{equation}
where \( G_{\ell} \) and \( \hat{G}_{\ell} \) are the true and predicted Legendre coefficients, respectively, and \( N_{\ell} \) is the number of terms in the expansion. To optimize computational efficiency and reduce overfitting, the expansion is truncated to 30 coefficients due to the rapid decay of higher-order coefficients.

In addition to the Legendre coefficients, the input feature vector includes a randomly generated Hubbard interaction parameter \( U \) and inverse temperature \( \beta \). By incorporating these parameters, the model captures their effects on the interacting Green’s function. This enables the network to learn a mapping from the non-interacting Green’s function, \( U \), and \( \beta \) to the Legendre coefficients of the interacting Green’s function, providing a compact and efficient prediction mechanism.

\subsection{Generating Random Band Structures} 
\begin{figure}
    \centering
    \includegraphics[width=\linewidth]{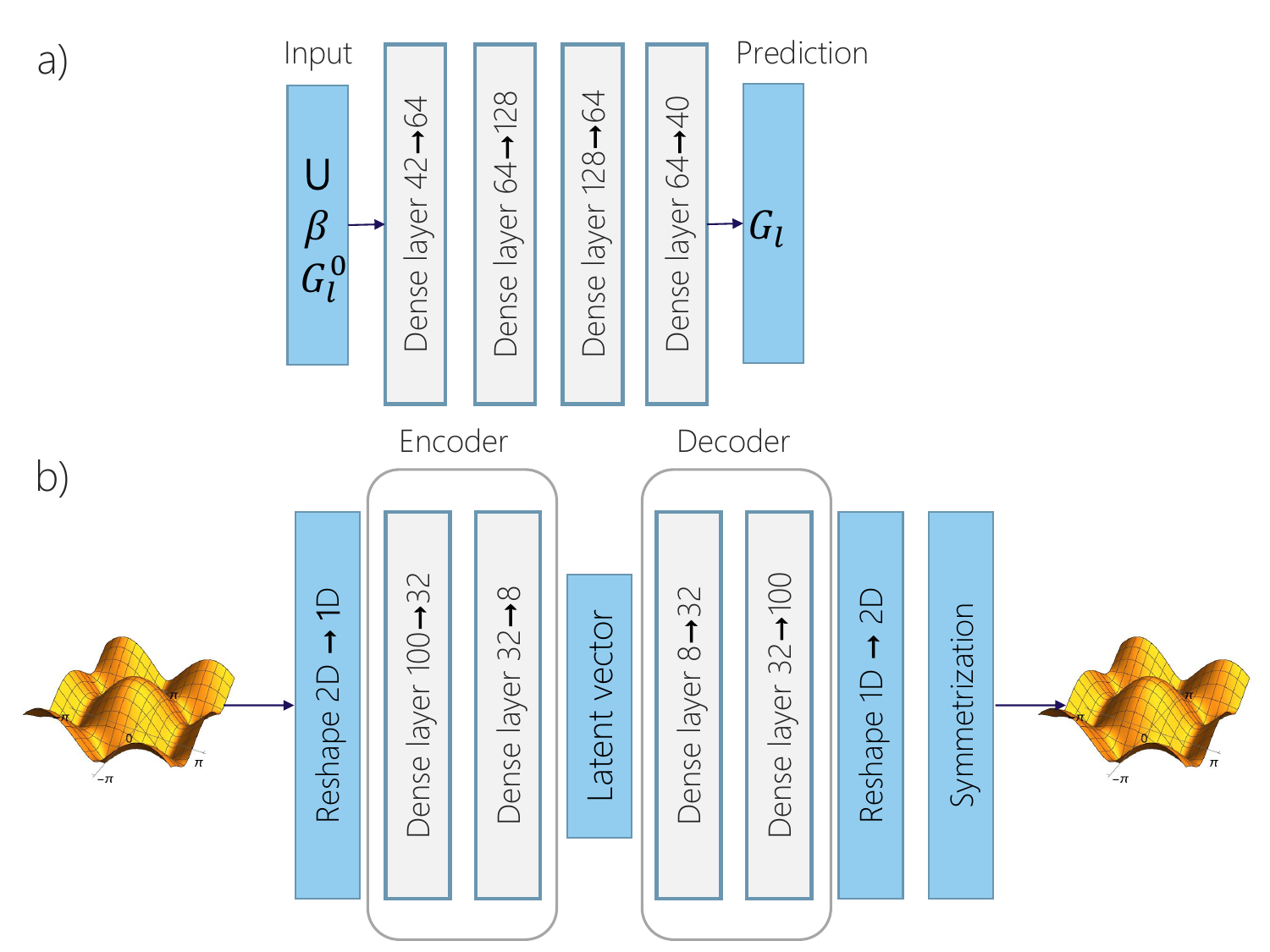}
    \caption{(a) Architecture of the primary neural network predicting the interacting Green's function \( G_l \) from the non-interacting \( G_l^0 \), the Hubbard interaction parameter \( U \), and the inverse temperature \( \beta \). 
    (b) Auxiliary neural network architecture used to generate random band structures.}
    \label{fig:arch}
\end{figure} 

\begin{figure}
    \centering
    \includegraphics[width=\linewidth]{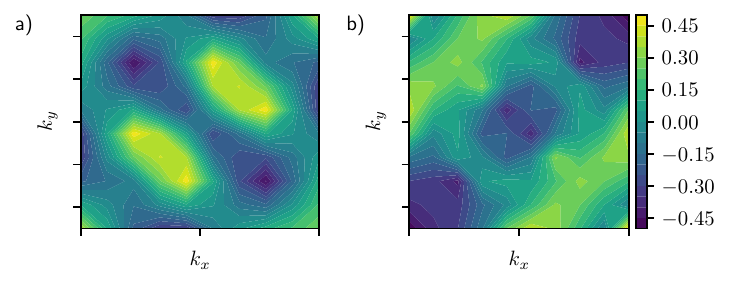}
    \includegraphics[width=\linewidth]{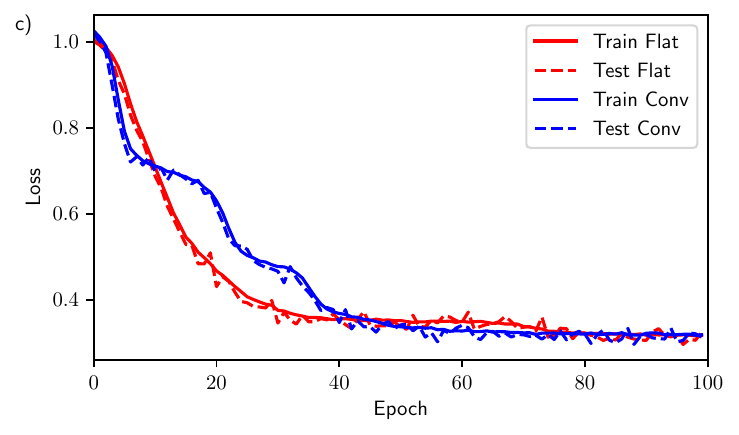}
    \caption{(a) Example of an extracted band structure from the database, (b) generated band structure, (c) learning process for different architectures.}
    \label{fig:auxtrain}
\end{figure}

To facilitate the generation of random band structures, we developed an auxiliary neural network using PyTorch~\cite{torch}. The autoencoder-based architecture, shown in \autoref{fig:arch} (b), is trained on band structures extracted from known materials (\autoref{fig:auxtrain} (a)). Once trained, the decoder serves as a generative model that produces new, physically plausible band structures from random, normally-sampled latent vectors (\autoref{fig:auxtrain} (b)). This approach enables scalability compared to directly using extracted band structures.

We sourced data from the Materials Project~\cite{jain_commentary_2013} and 2DMatPedia database~\cite{zhou_2dmatpedia_2019}, selecting over 6,000 materials with explicit rectangular \( \boldsymbol{k} \)-grid band structures. After filtering for bandwidths greater than 0.5 eV, over 1,000 materials with approximately 15,000 bands were retained. The bandwidth threshold ensures a focus on models with sparse density of states near the Fermi level. While our study emphasizes these systems, the inclusion of other band structures would be straightforward.

To preprocess the data, we standardized the \( \boldsymbol{k} \)-meshes by applying linear interpolation to a uniform 10-point grid along each dimension. The resulting two-dimensional arrays were flattened into one-dimensional vectors for input to the neural network. Although convolutional layers could encode the full two-dimensional structure, the flattening approach simplifies the model and suffices for our application, where precise spatial reconstruction is not required.

The autoencoder consists of an encoder, a latent layer, and a decoder that mirrors the encoder structure. The encoder features two fully connected layers with 100 and 32 neurons, respectively, followed by a latent layer with 8 neurons. The latent layer plays a critical role in controlling the sharpness of the band structures. The decoder mirrors the encoder with layers of 32 and 100 neurons, and a symmetrization step \( X + X^T \) is applied to the output along both diagonals to preserve the physical properties of the band structures.

The network minimizes the mean squared error (MSE) loss:
\begin{equation}
    \mathrm{MSE}=\frac{1}{S_{\text{BZ}} } \sum_{\boldsymbol{k} \in \text{BZ}} \left(\epsilon_{\boldsymbol{k}}-\hat{\epsilon}_{\boldsymbol{k}}\right)^2,
\end{equation}
where \( S_{\text{BZ}} \) is the Brillouin zone (BZ) volume, \( \epsilon_{\boldsymbol{k}} \) is the true energy band, and \( \hat{\epsilon}_{\boldsymbol{k}} \) is the predicted energy band. Training was performed using the Adam optimizer with a learning rate of \( 3 \times 10^{-3} \) over 100 epochs until convergence (\autoref{fig:auxtrain} (c)).

\subsection{Training the Primary Neural Network}

Using the auxiliary ML model, we generate random band structures to create a training set of interacting and non-interacting Green’s functions. The target outputs (interacting Green’s functions) are computed using TRIQS~\cite{parcollet_triqs_2015}, employing the continuous-time QMC interaction-expansion solver (CT-HYB)~\cite{seth_triqscthyb_2016} with density-density interactions. For the simple Hamiltonian used, the solver is configured to perform 4000 cycles of 1000 iterations, with 5000 warm-up cycles. All calculations include 1025 fermionic Matsubara frequencies.
This procedure yields a database of Green’s functions comprising 2000 rows, with inverse temperatures \( \beta \) ranging from 5 to 20 eV\(^{-1}\) and Hubbard interaction \( U \) values between 1.0 and 5.0 eV.

\begin{figure}
\centering
\includegraphics[width=\columnwidth]{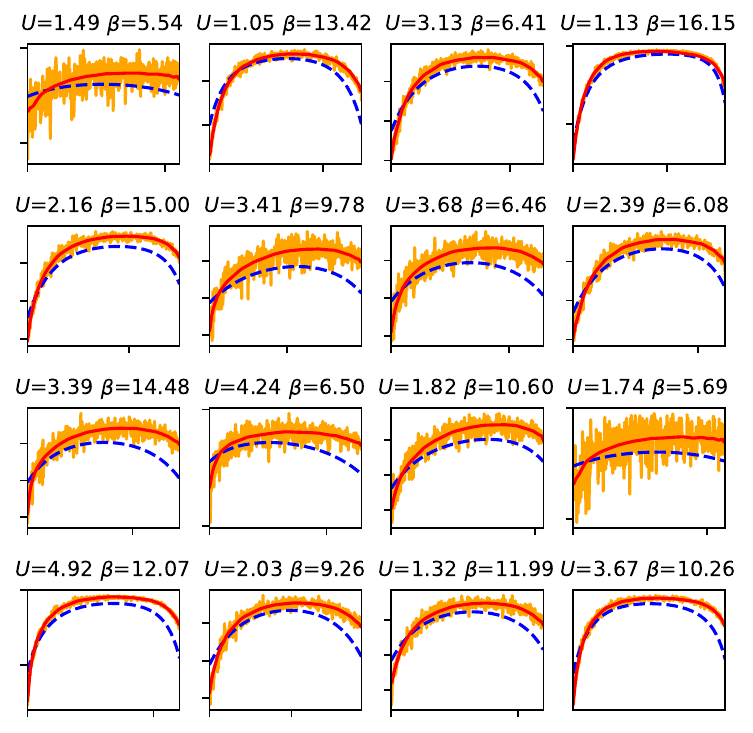}
\caption{Predictions of the ML model for Green’s functions in the \( \tau \)-basis. Dotted blue: non-interacting Green’s functions (\( G_0(\tau) \)), orange: interacting Green’s functions (\( G(\tau) \)), and solid red: ML-predicted Green’s functions (\( G_{\text{ML}}(\tau) \)). The horizontal axis is imaginary time (\( \tau \)) from 0 to \( \beta \), and the vertical axis is autoscaled. Interaction strength \( U \) is in eV, and \( \beta \) is in eV\(^{-1}\).}
\label{fig:results}
\end{figure}

The dataset is divided into training, validation, and test sets in a 70:20:10 ratio. The final mean-squared error (MSE) on the test set is approximately \( 1 \times 10^{-5} \), consistent with the training and validation losses. Predicted Green’s functions are shown in \autoref{fig:results}.

We obtained densities of states (DOS) from Green’s functions via analytic continuation using the maximum entropy method~\cite{jarrell_bayesian_1996,silver_maximum-entropy_1990} implemented in the Maxent package~\cite{kraberger_maximum_2017}. Compared to Padé continuation, this method avoids parameter tuning for individual samples. However, the inherent instability of analytic continuation leads to significant changes in the DOS for small errors in \( G(\tau) \).

\begin{figure}
    \centering
    \includegraphics[width=0.99\columnwidth]{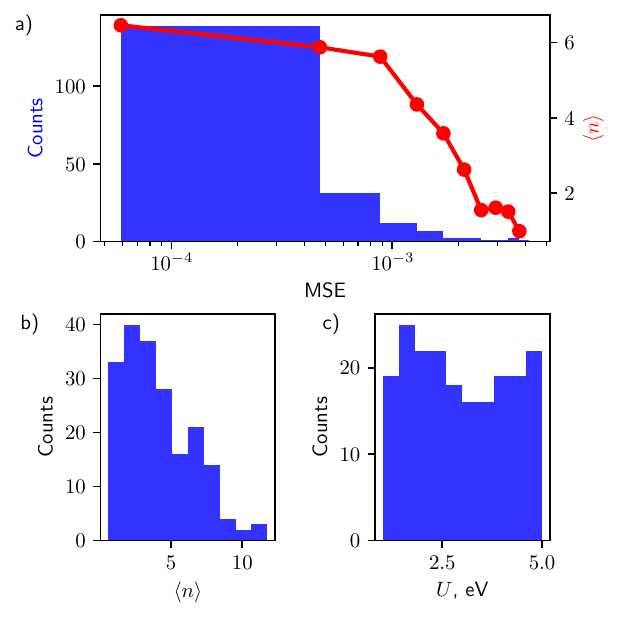}
    \caption{(a) Dispersion of the MSE on the test set, with the line showing the average perturbation order \( \langle n \rangle \). (b) Distribution of \( \langle n \rangle \) in the test set. (c) Distribution of \( U \) in the test set.}
    \label{fig:errors}
\end{figure}  

Interestingly, stronger interactions (\( U \)) do not result in larger errors, as shown in \autoref{fig:errors}. Lower errors are observed in regions with higher average perturbation orders \( K \), demonstrating the capability of the machine learning model to capture complex many-body physics, even in strongly correlated regimes. These trends are not merely due to the distribution of \( K \) in the training set, emphasizing the robustness of the predictions.

\section{Integrating the Solver into the DMFT Framework}

\begin{figure}
\centering
\includegraphics[width=0.99\linewidth]{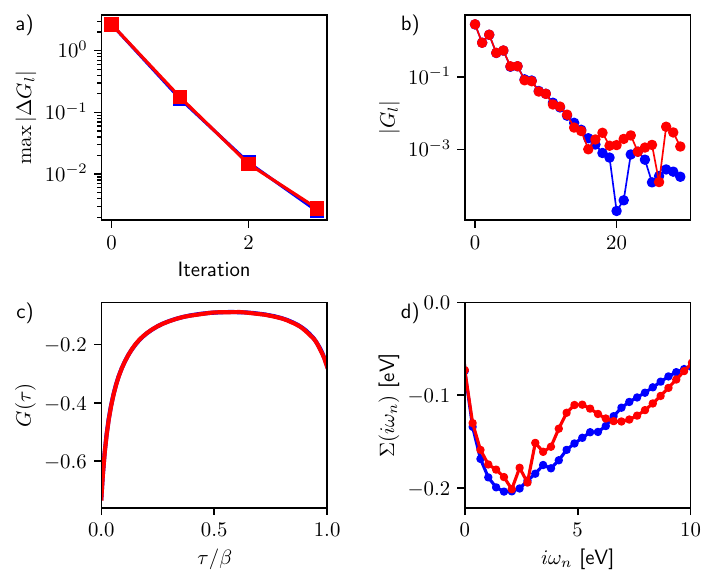}
\hspace{3 mm} 
\includegraphics[width=0.99\linewidth]{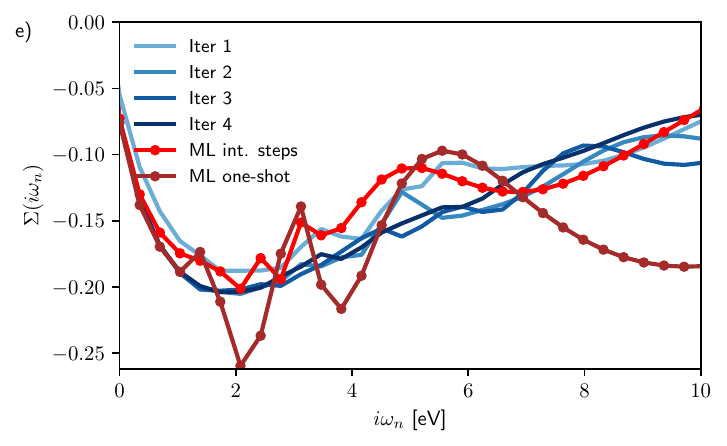}
\caption{This plot shows various properties of the exact DMFT (blue) and ML solution (red) with parameters $U=2.2$ eV and $\beta = 18.1$ eV$^{-1}$.  This includes:
(a) Convergence of iterations.
(b) Legendre polynomials.
(c) Imaginary time representation.
(d) Imaginary part of self-energy.
(e) Imaginary part of self-energy convergence and ML result. The ML one-shot stands for ML model trained on single-shot task,  ML int. steps is additionally trained on intermediate iterations.
}
\label{fig:DMFT_both}
\end{figure}

Once trained, the ML solver was integrated into the Dynamical Mean-Field Theory (DMFT) framework to accelerate the self-consistent calculations. For DMFT routines, we utilized the Two-Particle Response Function (TPRF) package of TRIQS~\cite{parcollet_triqs_2015}. Initial results showed that a single-shot task (where the model is trained only to predict the final interacting Green’s function) delivered suboptimal results compared to a model trained on a database incorporating intermediate iterations (\autoref{fig:DMFT_both} (e)).

To improve performance, we explored training the model to predict the self-energy directly in terms of Legendre polynomial coefficients. However, this approach was hindered by the slow decay of these coefficients and their constant bias in the frequency domain, which posed challenges for accurate predictions.

To address these issues, an additional database incorporating intermediate iterations was built by performing full DMFT calculations on 100 sample configurations. To ensure optimal spacing of measurements and reduce autocorrelation, we employed an adaptive cycle length strategy. This approach maintained the autocorrelation time between 0.7 and 1.5 by dynamically adjusting the cycle length: doubling it when the autocorrelation time exceeded 1.5 and halving it when it dropped below 0.7. This strategy stabilized measurements and improved data quality.

Each simulation was limited to 5 million measurement cycles, preceded by 3000 warm-up cycles, with an initial cycle length of 15. The Green’s functions were represented using 30 Legendre polynomials, 400 Matsubara frequencies (positive half-axis), and 4001 imaginary time points. Convergence was monitored using the difference between successive iterations of the Legendre polynomial coefficients, with a threshold of \( 5 \cdot 10^{-3} \), typically achieving convergence within 5 iterations.

To reduce noise at each iteration, the interacting Green’s function was transformed into the Legendre polynomial basis and then mapped back to Matsubara frequencies. While some samples exhibited difficulties in capturing high-frequency moments, the low-frequency self-energy generally maintained a reasonable structure.

The ML model previously trained on single-shot tasks was fine-tuned on this new data. This model demonstrated convergence rates comparable to the exact DMFT method (\autoref{fig:DMFT_both} (a)). However, while the ML solver achieved convergence, its final self-energy precision was slightly lower than that obtained after a few iterations of exact DMFT.

This approach shows promise for accelerating DMFT calculations by effectively skipping initial iterations. Reliable results currently require subsequent quantum Monte Carlo (QMC) iterations, but the number of iterations is reduced compared to standard DMFT (\autoref{fig:DMFT_both} (e)). Notably, even a single QMC iteration takes approximately 10 minutes on the same hardware, emphasizing the computational savings achieved.

\begin{figure}
\centering
\includegraphics[width=0.95\linewidth]{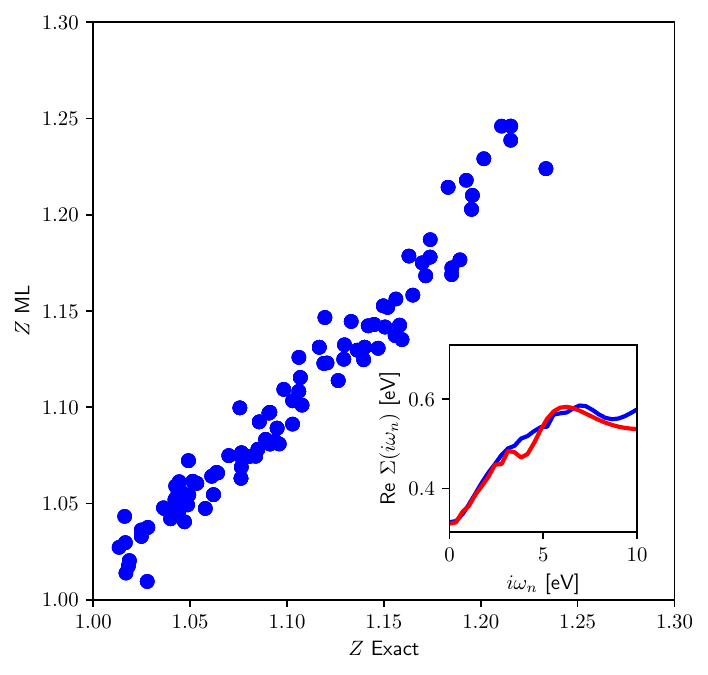}
\caption{Exact and predicted quasi-particle weights predicted from the ML model. Inset: real part of the self-energy for one sample. Blue: extracted from the QMC solution, red: extracted from the ML model.}
\label{fig:Z}
\end{figure}

The ML solver reliably predicts the quasi-particle weight, defined as:
\[
Z=\left(1-\left.\frac{\partial \operatorname{Re} \Sigma(i \omega)}{\partial i \omega}\right|_{\omega \rightarrow 0}\right)^{-1},
\]
which serves as an indicator of correlated behavior (values significantly larger than 1). This capability highlights the potential of the ML model for material screening (\autoref{fig:Z}).

\section{Conclusion}
In this work, we have demonstrated the feasibility of using machine learning to predict interacting Green’s functions on arbitrary 2D lattices in the single-orbital case. By leveraging dense neural networks and formulating the problem in the Legendre polynomial basis, we achieved accurate predictions of Green’s functions, highlighting the potential of machine learning for accelerating many-body physics simulations. Furthermore, we introduced an auxiliary neural network to generate random 2D band structures, enabling the creation of diverse training datasets that are scalable to more complex multi-band systems.

Our approach extends to integrating the machine learning solver within the Dynamical Mean-Field Theory (DMFT) framework. Fine-tuning the ML model on intermediate iterations allowed us to achieve convergence rates comparable to exact DMFT. While the self-energy predictions of the ML model were less precise than those obtained after a few iterations of exact DMFT, the solver effectively reduced the number of QMC iterations required, significantly accelerating the overall process. This hybrid approach positions the ML solver as a valuable tool for streamlining DMFT workflows.

An additional strength of the ML model is its ability to reliably predict the quasiparticle weight, \( Z \), a critical parameter for understanding correlated behavior in materials. This capability positions the model as a promising tool for screening materials based on their electronic properties, particularly for identifying candidates with strong correlations. However, full precision in self-energy and other key quantities still requires subsequent QMC iterations, emphasizing the complementary nature of the ML solver rather than its replacement of traditional methods.

Despite these advancements, several challenges remain. The lower accuracy in self-energy predictions, particularly in strongly correlated regimes, underscores the need for further refinement of the neural network architecture and training strategies. Future work could explore hybrid approaches, incorporating both ML predictions and QMC sampling, to achieve higher precision while retaining computational efficiency. Additionally, extending the method to multi-band systems, finite-temperature studies, and non-equilibrium dynamics presents exciting opportunities for advancing the utility of ML in DMFT.

In summary, this work establishes a foundational framework for integrating machine learning into DMFT, showcasing its potential to accelerate simulations while maintaining accuracy. The demonstrated ability to predict Green’s functions and quasiparticle weights suggests that ML-driven solvers could play a pivotal role in high-throughput material discovery, particularly in the search for novel correlated materials.

\section*{Acknowledgements} 
We are grateful A. A. Katanin for useful discussions of the DMFT algorithm and A.~A.~Khomutov for discussions regarding ML model architectures.
This project started as part of the Quantum Entrepreneurship Lab course at the Technical University of Munich (TUM), led by a collaboration between PushQuantum e.V. and TUM Venture Labs Quantum.
This research is part of the Munich Quantum Valley, which is supported by the Bavarian state government with funds from the Hightech Agenda Bayern Plus.

\section*{Code Availability}
Codes for \href{https://github.com/EgorcaA/band_gen}{Band Generation},  \href{https://github.com/EgorcaA/gf2ML}{Green's function prediction}, and 
\href{https://github.com/EgorcaA/DMFT_ML}{DMFT implementation}  
are available in their linked GitHub repositories.
The rest of the data is available upon request.

\bibliography{bibl.bib}

\end{document}